\documentclass[prl,twocolumn,a4paper]{revtex4}

\usepackage{amssymb}
\usepackage{graphicx}
\usepackage{verbatim}
\usepackage{eucal}
\usepackage{epstopdf}

\newcommand{\ket}[1]{\ensuremath{\left| #1 \right\rangle}}

\begin{document}

\title{High-fidelity transmission of entanglement over a high-loss freespace channel}

\author{Alessandro Fedrizzi$^{1}$, Rupert Ursin$^1$, Thomas Herbst$^1$, Matteo Nespoli$^1$, Robert Prevedel$^1$, Thomas Scheidl$^1$, Felix Tiefenbacher$^1$, Thomas Jennewein$^1$,\\and Anton Zeilinger$^{1,2}$}
\email{zeilinger-office@exp.univie.ac.at}

\affiliation{$^1$Institute for Quantum Optics and Quantum
Information, Austrian Academy of Sciences, Boltzmanngasse 3, 1090
Wien, Austria
 \\$^2$Quantum Optics, Quantum Nanophysics and Quantum Information, Faculty of Physics, University of Vienna,
    Boltzmanngasse 5, 1090 Vienna, Austria}
\date{\today}

\maketitle
\textbf{Quantum entanglement enables tasks not possible in classical physics. Many quantum communication protocols \cite{gisin2007qc} require the distribution of entangled states between distant parties. Here we experimentally demonstrate the successful transmission of an entangled photon pair over a 144 km free-space link. The received entangled states have excellent, noise-limited fidelity, even though they are exposed to extreme attenuation dominated by turbulent atmospheric effects. The total channel loss of 64 dB corresponds to the estimated attenuation regime for a two-photon satellite quantum communication scenario. We confirm that the received two-photon states are still highly entangled by violating the CHSH inequality by more than 5 standard deviations. From a fundamental point of view, our results show that the photons are subject to virtually no decoherence during their 0.5 ms long flight through air, which is encouraging for future world-wide quantum communication scenarios.}

Entanglement is at the heart of many peculiarities encountered in quantum mechanics and has allowed many ground-breaking tests on the fundamentals of nature. Entangled photons  are ideal tools to investigate the laws of quantum mechanics over long distances and time-scales since they are not subject to decoherence. Furthermore photons can be easily generated, manipulated and transmitted over large distances via optical fibres or free-space links. Since the maximal distance for the distribution of quantum entanglement in optical fibres is limited to the order \cite{takesue2007qkd,hubel2007hft,honjo2007ldd,zhang2007dte} of $\sim100$ km, the most promising option for testing quantum entanglement on a global scale is currently free-space transmission, ultimately using satellites and ground stations \cite{aspelmeyer2003ldq}. 

In recent years, various free-space quantum communication experiments with weak coherent laser pulses \cite{Kurtsiefer2002,buttler2000dqk,Rarity2001,bienfang2004qkd,Schmitt-Manderbach2007} and entangled photons \cite{aspelmeyer2003ldf,peng2005efs,resch2005dea,marcikic2006fsq} have been performed on ever larger distance scales and with increasing bit rates. The to-date most advanced test bed for free-space distribution of entanglement is a 144 km free-space link between two Canary Islands, where the successful transmission of one photon of an entangled pair was recently achieved \cite{ursin2007ebq}.  In the present experiment we demonstrate a fundamentally more interesting scenario by sending both photons of an entangled pair over this free-space channel. By violating a Clauser-Horne-Shimony-Holt (CHSH) Bell inequality \cite{clauser1969pet} we find that entanglement is highly stable over these long time spans - the photon-pair flight time of $\sim0.5~$ms is the longest lifetime of photonic Bell states reported so far, almost twice as long as the previous high \cite{honjo2007ldd,zhang2007dte} of $\sim250~\mu$s.

The achieved noise-limited fidelity paves the way for free-space implementations of quantum communication protocols that require the transmission of two photons, e.g. quantum dense coding \cite{mattle1996}, entanglement purification \cite{pan2003eep}, quantum teleportation \cite{bouwmeester2001pqi} and quantum key distribution without a shared reference frame \cite{boileau2004rpb}. From a technological perspective, the overall two-photon loss bridged in our experiment is significantly higher than the current 40 dB limit \cite{ma2007qkd} for alternative setups relying on weak coherent laser pulses. The attenuation of 64 dB corresponds to the expected attenuation for a satellite scenario with two ground stations \cite{aspelmeyer2003ldq}, proving the feasibility of quantum communication on a global scale.

The experiment was conducted between La Palma and Tenerife, two Canary islands situated in the Atlantic ocean off the West African coast. 
\begin{figure*}[!htbp]
\centering\includegraphics[width=.9\textwidth]{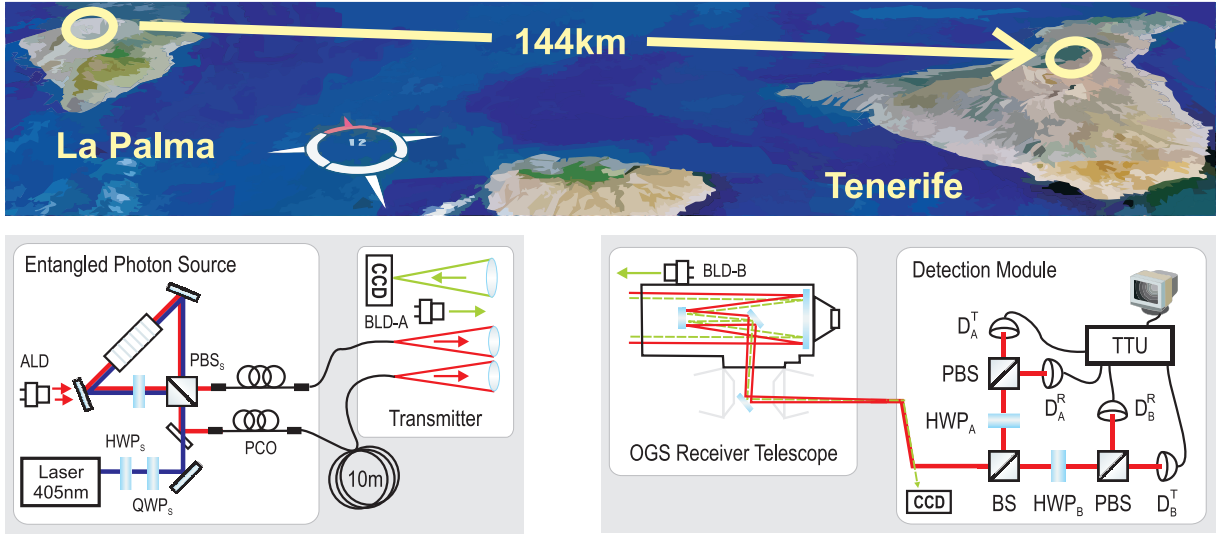}
\caption{Satellite image (NASA World Wind) of the Canary islands of Tenerife and La Palma and overview of the experimental scheme. At La Palma, a Sagnac down-conversion source created narrow-band entangled photon pairs. Manual polarization controllers (PCO) and an auxiliary laser diode (ALD) were used for polarization alignment. The photon pairs were transmitted via a pair of telescopes mounted on a rotatable platform to the 144 km distant receiver (OGS) on Tenerife. The transmitter telescope platform was actively locked to a $532$~nm beacon laser attached to the OGS (BLD-B). Likewise, the OGS receiver telescope tracked the virtual position of a $532$~nm beacon laser attached to the transmitter (BLD-A). At the OGS, the overlapping photon beams were collected and guided to the detection module by a system of mirrors. This module consisted of a 50/50 beamsplitter cube (BS) and two polarization analyzers (A,B). Each of these analyzers was formed by one half-wave plate (HWP$_A$, HWP$_B$), a polarizing beamsplitter cubes (PBS) and two single-photon silicon avalanche photo diodes (D$_A^T$, D$_A^R$, D$_B^T$, D$_B^R$) placed in the transmitted (T) and the reflected (R) output port of the respective PBS.} \label{fig:setup}
\end{figure*}
An overview of the experimental scheme is shown in figure \ref{fig:setup}. At the transmitter station at La Palma, photon pairs at a wavelength of $810$~nm and a bandwidth (FWHM) of $0.6$~nm were generated in a $10$~mm long, periodically poled KTiOPO$_4$ crystal which was bidirectionally pumped by a grating-stabilized $405$~nm diode laser. The photon pairs were coherently combined in a polarization Sagnac interferometer \cite{kim2006pss,fedrizzi2007wtf} and emitted in the maximally entangled state:
\begin{equation}\label{eq:psiminus}
\frac{1}{\sqrt{2}}\ket{HV}+e^{i\varphi}\ket{VH}.
\end{equation}
At 20mW of pump power, the source produced $\sim10^7$ photon pairs/s of which $\sim3.3\times10^6$ single photons/s and $\sim10^6$ pairs/s were detected locally. These pairs were coupled into single-mode fibres with a length difference of $10$~m which introduced a time delay of $\Delta t=50$~ns between the two photons. The two photons were transmitted by two telescopes mounted on a motorized platform to a common receiver telescope Ð the European Space AgencyÕs Optical Ground Station (OGS) located on Tenerife. The transmitters consisted of single-mode fibre couplers and f/4 best form lenses (focal length $f=280$~mm) which had a lateral separation of $10$~cm. To actively compensate the transmitter platform and receiver pointing directions for drifts of the optical path through the time-dependent atmosphere, a bidirectional closed-loop tracking mechanism was employed: At the transmitter platform, the virtual position of a $532$~nm beacon laser attached to the OGS was monitored in the focus of a third telescope by a CCD camera. Likewise, the OGS monitored the position of a beacon laser mounted at the transmitter (see figure (1) and \cite{Schmitt-Manderbach2007} for details). 

\begin{figure*}[!htbp]
\centering\includegraphics[width=0.6\textwidth]{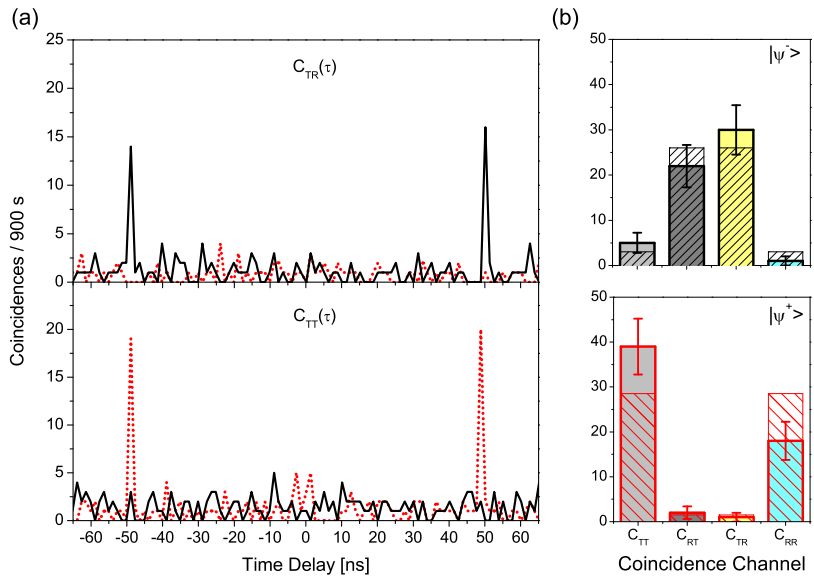}
\caption{Coincidence histograms and the respective accumulated coincidence events for measurements on two different Bell states. (a) Timing distribution of two out of four coincidence channels. C$_{TR}(\tau)$ and C$_{TT}(\tau)$ between detectors D$_A^T$-D$_B^R$ (top) and D$_A^T$-D$_B^T$ (bottom) for a $\ket{\Psi^-}$ state (black line) and a $\ket{\Psi^+}$ state (dotted red line). The analyzer wave plates HWP$_A$ and HWP$_B$ in the detection module were set to ($\pi/8$, $\pi/8$). For each detector combination, there are two coincidence peaks at $\pm50$~ns which can be clearly distinguished from the accidental background. (b) Total coincidence counts and Poissonian standard deviations for all four relevant coincidence channels integrated over a $1.25$~ns time window centred at the peak positions. They show distinct $\ket{\Psi^-}$ (top) and $\ket{\Psi^+}$ (bottom) signatures. The full-colored columns show the joint results of the four corresponding detector combinations necessary to fully characterize the state. The hatched columns show the ideal expectation. The accumulation time for each measurement was 900 seconds.} \label{fig:coincidences}
\end{figure*}

At the OGS, the incoming photons were collected by a 1 m mirror ($f=38$~m). To ensure that turbulence induced beam wander did not divert the beam off the detectors, the photons were collimated ($f=400$~mm) before they were guided to a polarization analysis module. Here, a symmetric $50/50$ beam splitter (BS) directed impinging photons randomly to one of two polarization analyzers (A, B), each consisting of a half-wave plate (HWP$_A$, HWP$_B$), mounted in a motorized rotation stage, and a polarizing beam splitter (PBS). The polarized light was then refocused ($f=50$~mm) onto four single-photon avalanche diodes (SPADs). A time-stamping unit recorded clicks in the four SPADs and encoded and stored their respective channel number and arrival time relative to a common internal clock with 156 ps resolution. Figure \ref{fig:coincidences} (a) shows the cross-correlations of these time-stamps for two exemplary measurements. One can clearly identify the two coincidence peaks at $\pm50$~ns around zero delay, which corresponds to the fibre delay $\Delta t$ introduced at the transmitter. The average width (FWHM) of the coincidence peaks was $560$~ps, dominated by the timing-jitter of the SPADs. To obtain the number of coincident photons, we summed up the number of correlations in a time window of $1.25$~ns centred at the coincidence peak positions (figure \ref{fig:coincidences} (b)). 

As a witness for the presence of entanglement between the received photons, we tested the CHSH Bell inequality \cite{clauser1969pet}:
\begin{eqnarray}\label{eq:chsh}
S(\alpha,\beta,\alpha',\beta')=\mid
E(\alpha,\beta)-E(\alpha,\beta')\mid+\nonumber\\
+\mid E(\alpha',\beta')+E(\alpha',\beta)\mid\leq2,
\end{eqnarray}
where $E(\alpha,\beta)=(C_{TT}(\theta_A,\theta_B)+C_{RR}(\theta_A,\theta_B)-C_{TR}(\theta_A,\theta_B)-C_{RT}(\theta_A,\theta_B))/N$ is the normalized correlation value of polarization measurement results on photon pairs. $C_{ab}(\theta_A,\theta_B)$ is the number of coincidences measured between detectors at the (T/R) output ports (figure(1)) of the polarization analysers $A$ and $B$ set to angles $(\theta_A,\theta_B)$ and $N$ is the sum of these coincidences. Whenever $S$ exceeds the classical bound $S>2$, the polarization correlations cannot be explained by local hidden-variable models \cite{clauser1969pet}. For the maximally entangled state $\ket{\Psi^-}$, quantum theory predicts a limit of $S_{\rm{QM}}=2\sqrt{2}$ for the settings $(\alpha,\beta,\alpha',\beta')=(0,\pi/8,\pi/4,3\pi/8)$.

The detection module in our experiment allowed us to directly measure the expectation values $E(\theta_A, \theta_B)$  in equation \ref{eq:chsh} with 4 different sets of angles of HWP$_A$ and HWP$_B$ (table \ref{tab:chshresults}). We first aligned the system to obtain a $\ket{\Psi^-}$ state at the receiver (details in the Methods section). For each setting $(\theta_A, \theta_B)$ we repeatedly accumulated data for typically $900$ seconds, which eventually amounted to a total of $10800$ seconds acquired in three consecutive nights. Each detector registered an intrinsic dark count rate of $\sim200/s$, and additionally background light of $\sim200/s$. In total, we received an average signal of $2500$ single-photons/s and $0.071$ photon pairs/s. Even though the final single-photon-to-coincidence ratio at the receiver was just $1.7\times10^{-5}$, the coincidence signal-to-noise ratio (SNR) was as high as $\sim15:1$. Compared to the count rates detected at the source, the single-photon attenuation was $34$~dB, of which $2$~dB were due to the lower efficiency of the detectors used at the receiver ($\sim25\%$) compared to those at the source ($\sim40\%$). The measured, total photon-pair loss was $71~dB$, of which $3~dB$ were contributed by the BS in the receiver module. The average net attenuation experienced by single photons along the free-space link was therefore $32~dB$ and the photon-pair attenuation of $64$~dB was exactly twice as large, which results from the fact that a pair of photons has double the extinction of a single photon transiting the path. As we have previously ruled out any other adverse effects such as depolarization or timing jitter, which might occur between photons in independent channels \cite{ursin2007ebq}, we can compare our results to a scenario with two separate free-space links. A detailed analysis \cite{ma2007qkd} of the error sources in our system allowed us to estimate the expected background and multi-photon pair emissions limited quantum visibility to $94.4\%$. Combined  with the source visibility ($99.2\%$) and the polarization contrast of the detection module ($99.5\%$), the upper bound for the overall system visibility was $V_{\rm{tot}}=93.2\%$. As the observed CHSH Bell parameter is limited by the setup visibility via $S_{\rm{max}}=V_{\rm{tot}}\times S_{\rm{QM}}$,  this implies a maximum achievable Bell parameter of $S_{\rm{max}}=2.636$.

The accumulated coincident events for the different detector pairs yielded the correlation values shown in table \ref{tab:chshresults}. 
\begin{table} [!hbpt]
\begin{tabular}{ | c | c | c | }
\hline
$(\theta_a,\theta_b)$ & E$(\theta_a,\theta_b)$ & $\Delta$E$(\theta_a,\theta_b)$ \\
\hline
$(0,\pi/8)$ & -0.604 & 0.059 \\
$(\pi/4,\pi/8)$ & 0.672 & 0.055 \\
$(0,3\pi/8)$ & 0.638 & 0.056 \\
$(\pi/4,3\pi/8)$ & 0.697 & 0.058 \\
\hline
\end{tabular}
\caption{Experimental polarization correlations E$(\theta_a,\theta_b)$ for the CHSH inequality. the total integration time was 10800 seconds. The standard deviations $\Delta$E$(\theta_a,\theta_b)$ were calculated assuming Poissonian photon count statistics.}
\label{tab:chshresults}
\end{table}

According to equation \ref{eq:chsh}, we measured a CHSH Bell parameter $S_{\rm{exp}}$ of:
\begin{equation}\label{eq:svalue}
S=2.612\pm0.114.
\end{equation}
which is in excellent agreement to our estimate $S_{\rm{max}}$. Our result violates the CHSH inequality by $5.4$ standard deviations and convincingly proves the successful transmission of entanglement. The fact that $S_{\rm{exp}}$ is so close to $S_{\rm{max}}$ shows that the fidelity between the transmitted and received entangled states was essentially noise-limited. Therefore, the entanglement was not affected by decoherence, event though the photons were subject to extreme attenuation which was dominated bu turbulent atospheric fluctuation \cite{ursin2007ebq}.

Note that our setup contains some of the basic building blocks of a quantum communication system. However, the setup could not be used to carry out an actual quantum key distribution experiment \cite{bennett1992qcw} owing to the lack of a second independent analyzer module. In addition, a full-fledged implementation would have required classical post-processing protocols, i.e. error correction and privacy amplification. Nevertheless, from the measured $S_{\rm{exp}}$, we can infer a qubit error ratio of $3,85\pm2,2\%$. 

Further note that the locally detected photon-pair rate of $\sim10^6$ pairs/s at the source is necessary for the long-distance distribution of quantum entanglement in the high-attenuation regime \cite{aspelmeyer2003ldq}. The compact entangled-photon source, being pumped by a low-power diode laser, can readily be integrated into a satellite-borne photonic terminal, which was previously\cite{resch2005dea,ursin2007ebq} not the case. We expect that this will allow fundamental tests of the laws of quantum mechanics on a global scale \cite{perdigues2007qce}.

\subsection{Methods}
Before measuring the polarization correlations for equation \ref{eq:chsh}, we had to establish a common polarization reference frame between the individual transmitters and the receiver and to adjust the phase $\varphi$ of the quantum state (\ref{eq:psiminus}) such that the detected coincidence signature was consistent with one of the desired Bell states. For the polarization compensation, we used an auxiliary 808 nm laser diode, which was directed at  the entangled photon source such that linearly polarized light was coupled into the fibres at a well defined single photon level (figure \ref{fig:setup}). We set HWP$_A$ and HWP$_B$ in the detection module to ($0\symbol{23}$, $0\symbol{23}$), measuring in the $\ket{H/V}$ basis and manually adjusted the fibre polarization controllers to at the source to maximize the single-photon polarization visibility $V_{H/V}$ in the remote detectors. The achieved visibility in this basis was typically $95\%$.

\begin{figure}[!htbp]
\centering\includegraphics[width=0.4\textwidth]{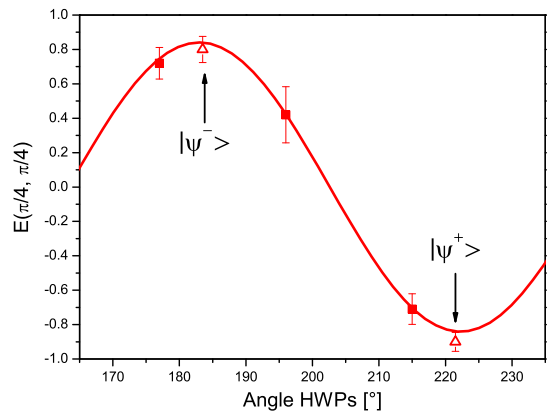}
\caption{Scan of the phase $\varphi$ of the entangled two-photon state in one measurement night. We measured the visibility of the entangled states in the $\ket{\pm}$ basis for three settings of the wave plates controlling the pump laser polarization (squares). After fitting a cosine function ($V_0=84\pm2.4\%$) to these data points, we were able to adjust the source to emit either $\ket{\Psi^-}$ or $\ket{\Psi^+}$ states. We prepared these states and observed, in this example, a visibility of $V_\pm=80\pm7.6\%$ for the $\ket{\Psi^-}$ and $V_\pm=90\pm5.5\%$ for the $\ket{\Psi^-}$ state (triangles). The error bars were determined using Poissonian photon count statistics.}\label{fig:phase}
\end{figure}

Once the linear polarization was set, the auxiliary laser diode was switched off and  $\varphi$ was adjusted using entangled photons. The visibility $V_{\pm}$ in the $\ket{\pm}=\ket{H\pm
V}$ basis depends on $\varphi$ as $V_{\pm}=V_0cos(\varphi)$. To determine the relation between $\varphi$ and the waveplates controlling the pump laser in the source (HWP$_{\rm{s}}$, QWP$_{\rm{s}}$), we set HWP$_A$ and HWP$_B$ in the detection module to ($\pi/8$, $\pi/8$), measured $V_{\pm}$ for three different settings of HWP$_{\rm{s}}$ and QWP$_{\rm{s}}$ in the equatorial plane of the Poincar\'e sphere representing the pump laser polarization (see \cite{kim2006pss,fedrizzi2007wtf}) and then numerically fitted a cosine function to the obtained data points. From this fit we deduced the waveplate settings to obtain either a $\ket{\Psi^-}$ or a $\ket{\Psi^+}$ state in the detection module. This procedure was repeated each night at the beginning of a measurement. The coincidence measurements in figure \ref{fig:coincidences} were obtained after we prepared $\ket{\Psi^-}$ or a $\ket{\Psi^+}$ states with this method. The measured respective visibility of $V_\pm=80\pm7.6\%$ and $V_\pm=90\pm5.5\%$ is shown in figure \ref{fig:phase}. As the linear polarization of the individual photons could be adjusted arbitrarily at the source, we were thus able to prepare, transmit and distinguish any of the 4 Bell states at the receiver.

\setcounter{secnumdepth}{0}

\subsection{Acknowledgments}
We are grateful to H. Weinfurter, J. G. Rarity, T. Schmitt-Manderbach, C. Barbieri, F. Sanchez, A. Alonso, J. Perdigues and Z. Sodnik, T. Augusteijn and the staff of the Nordic Optical Telescope in La Palma for their support at the trial sites. This work was supported by ESA under the General Studies Programme (No. 18805/04/NL/HE), the European Commission through Project QAP (No. 015846), the DTO-Funded U.S. Army Research Office, the Austrian Science Foundation (FWF) under project number SFB1520 and the ASAP-Programme of the Austrian Space Agency (FFG).


\begin{thebibliography}{26}
\expandafter\ifx\csname natexlab\endcsname\relax\def\natexlab#1{#1}\fi
\expandafter\ifx\csname bibnamefont\endcsname\relax
  \def\bibnamefont#1{#1}\fi
\expandafter\ifx\csname bibfnamefont\endcsname\relax
  \def\bibfnamefont#1{#1}\fi
\expandafter\ifx\csname citenamefont\endcsname\relax
  \def\citenamefont#1{#1}\fi
\expandafter\ifx\csname url\endcsname\relax
  \def\url#1{\texttt{#1}}\fi
\expandafter\ifx\csname urlprefix\endcsname\relax\def\urlprefix{URL }\fi
\providecommand{\bibinfo}[2]{#2}
\providecommand{\eprint}[2][]{\url{#2}}

\bibitem[{\citenamefont{Gisin and Thew}(2007)}]{gisin2007qc}
\bibinfo{author}{\bibfnamefont{N.}~\bibnamefont{Gisin}} \bibnamefont{and}
  \bibinfo{author}{\bibfnamefont{R.}~\bibnamefont{Thew}}, \bibinfo{journal}{Nat
  Photon} \textbf{\bibinfo{volume}{1}}, \bibinfo{pages}{165}
  (\bibinfo{year}{2007}).

\bibitem[{\citenamefont{Takesue et~al.}(2007)\citenamefont{Takesue, Nam, Zhang,
  Hadfield, Honjo, Tamaki, and Yamamoto}}]{takesue2007qkd}
\bibinfo{author}{\bibfnamefont{H.}~\bibnamefont{Takesue}},
  \bibinfo{author}{\bibfnamefont{S.~W.} \bibnamefont{Nam}},
  \bibinfo{author}{\bibfnamefont{Q.}~\bibnamefont{Zhang}},
  \bibinfo{author}{\bibfnamefont{R.~H.} \bibnamefont{Hadfield}},
  \bibinfo{author}{\bibfnamefont{T.}~\bibnamefont{Honjo}},
  \bibinfo{author}{\bibfnamefont{K.}~\bibnamefont{Tamaki}}, \bibnamefont{and}
  \bibinfo{author}{\bibfnamefont{Y.}~\bibnamefont{Yamamoto}},
  \bibinfo{journal}{Nature Photonics} \textbf{\bibinfo{volume}{1}},
  \bibinfo{pages}{343} (\bibinfo{year}{2007}).

\bibitem[{\citenamefont{H{\"u}bel et~al.}(2007)\citenamefont{H{\"u}bel, Vanner,
  Lederer, Blauensteiner, Lor{\"u}nser, Poppe, and Zeilinger}}]{hubel2007hft}
\bibinfo{author}{\bibfnamefont{H.}~\bibnamefont{H{\"u}bel}},
  \bibinfo{author}{\bibfnamefont{M.}~\bibnamefont{Vanner}},
  \bibinfo{author}{\bibfnamefont{T.}~\bibnamefont{Lederer}},
  \bibinfo{author}{\bibfnamefont{B.}~\bibnamefont{Blauensteiner}},
  \bibinfo{author}{\bibfnamefont{T.}~\bibnamefont{Lor{\"u}nser}},
  \bibinfo{author}{\bibfnamefont{A.}~\bibnamefont{Poppe}}, \bibnamefont{and}
  \bibinfo{author}{\bibfnamefont{A.}~\bibnamefont{Zeilinger}},
  \bibinfo{journal}{Optics Express} \textbf{\bibinfo{volume}{15}},
  \bibinfo{pages}{7853} (\bibinfo{year}{2007}).

\bibitem[{\citenamefont{Honjo et~al.}(2007)\citenamefont{Honjo, Takesue,
  Kamada, Nishida, Tadanaga, Asobe, and Inoue}}]{honjo2007ldd}
\bibinfo{author}{\bibfnamefont{T.}~\bibnamefont{Honjo}},
  \bibinfo{author}{\bibfnamefont{H.}~\bibnamefont{Takesue}},
  \bibinfo{author}{\bibfnamefont{H.}~\bibnamefont{Kamada}},
  \bibinfo{author}{\bibfnamefont{Y.}~\bibnamefont{Nishida}},
  \bibinfo{author}{\bibfnamefont{O.}~\bibnamefont{Tadanaga}},
  \bibinfo{author}{\bibfnamefont{M.}~\bibnamefont{Asobe}}, \bibnamefont{and}
  \bibinfo{author}{\bibfnamefont{K.}~\bibnamefont{Inoue}},
  \bibinfo{journal}{Optics Express} \textbf{\bibinfo{volume}{15}},
  \bibinfo{pages}{13957} (\bibinfo{year}{2007}).

\bibitem[{\citenamefont{Zhang et~al.}(2008)\citenamefont{Zhang, Takesue, Nam,
  Langrock, Xie, Fejer, and Yamamoto}}]{zhang2007dte}
\bibinfo{author}{\bibfnamefont{Q.}~\bibnamefont{Zhang}},
  \bibinfo{author}{\bibfnamefont{H.}~\bibnamefont{Takesue}},
  \bibinfo{author}{\bibfnamefont{S.~W.} \bibnamefont{Nam}},
  \bibinfo{author}{\bibfnamefont{C.}~\bibnamefont{Langrock}},
  \bibinfo{author}{\bibfnamefont{X.}~\bibnamefont{Xie}},
  \bibinfo{author}{\bibfnamefont{M.~M.} \bibnamefont{Fejer}}, \bibnamefont{and}
  \bibinfo{author}{\bibfnamefont{Y.}~\bibnamefont{Yamamoto}},
  \bibinfo{journal}{Optics Express} \textbf{\bibinfo{volume}{16}},
  \bibinfo{pages}{5776} (\bibinfo{year}{2008}).

\bibitem[{\citenamefont{Aspelmeyer
  et~al.}(2003{\natexlab{a}})\citenamefont{Aspelmeyer, Jennewein, Pfennigbauer,
  Leeb, and Zeilinger}}]{aspelmeyer2003ldq}
\bibinfo{author}{\bibfnamefont{M.}~\bibnamefont{Aspelmeyer}},
  \bibinfo{author}{\bibfnamefont{T.}~\bibnamefont{Jennewein}},
  \bibinfo{author}{\bibfnamefont{M.}~\bibnamefont{Pfennigbauer}},
  \bibinfo{author}{\bibfnamefont{W.}~\bibnamefont{Leeb}}, \bibnamefont{and}
  \bibinfo{author}{\bibfnamefont{A.}~\bibnamefont{Zeilinger}},
  \bibinfo{journal}{Selected Topics in Quantum Electronics, IEEE Journal of}
  \textbf{\bibinfo{volume}{9}}, \bibinfo{pages}{1541}
  (\bibinfo{year}{2003}{\natexlab{a}}).

\bibitem[{\citenamefont{Kurtsiefer et~al.}(2002)\citenamefont{Kurtsiefer,
  Zarda, Halder, Weinfurter, Gorman, Tapster, and Rarity}}]{Kurtsiefer2002}
\bibinfo{author}{\bibfnamefont{C.}~\bibnamefont{Kurtsiefer}},
  \bibinfo{author}{\bibfnamefont{P.}~\bibnamefont{Zarda}},
  \bibinfo{author}{\bibfnamefont{M.}~\bibnamefont{Halder}},
  \bibinfo{author}{\bibfnamefont{H.}~\bibnamefont{Weinfurter}},
  \bibinfo{author}{\bibfnamefont{P.}~\bibnamefont{Gorman}},
  \bibinfo{author}{\bibfnamefont{P.}~\bibnamefont{Tapster}}, \bibnamefont{and}
  \bibinfo{author}{\bibfnamefont{J.}~\bibnamefont{Rarity}},
  \bibinfo{journal}{Nature} \textbf{\bibinfo{volume}{419}},
  \bibinfo{pages}{450} (\bibinfo{year}{2002}).

\bibitem[{\citenamefont{Buttler et~al.}(2000)\citenamefont{Buttler, Hughes,
  Lamoreaux, Morgan, Nordholt, and Peterson}}]{buttler2000dqk}
\bibinfo{author}{\bibfnamefont{W.}~\bibnamefont{Buttler}},
  \bibinfo{author}{\bibfnamefont{R.}~\bibnamefont{Hughes}},
  \bibinfo{author}{\bibfnamefont{S.}~\bibnamefont{Lamoreaux}},
  \bibinfo{author}{\bibfnamefont{G.}~\bibnamefont{Morgan}},
  \bibinfo{author}{\bibfnamefont{J.}~\bibnamefont{Nordholt}}, \bibnamefont{and}
  \bibinfo{author}{\bibfnamefont{C.}~\bibnamefont{Peterson}},
  \bibinfo{journal}{Physical Review Letters} \textbf{\bibinfo{volume}{84}},
  \bibinfo{pages}{5652} (\bibinfo{year}{2000}).

\bibitem[{\citenamefont{Rarity et~al.}(2001)\citenamefont{Rarity, Tapster, and
  Gorman}}]{Rarity2001}
\bibinfo{author}{\bibfnamefont{J.~G.} \bibnamefont{Rarity}},
  \bibinfo{author}{\bibfnamefont{P.~R.} \bibnamefont{Tapster}},
  \bibnamefont{and} \bibinfo{author}{\bibfnamefont{P.~M.}
  \bibnamefont{Gorman}}, \bibinfo{journal}{Journal of Modern Optics}
  \textbf{\bibinfo{volume}{48}}, \bibinfo{pages}{1887} (\bibinfo{year}{2001}).

\bibitem[{\citenamefont{Bienfang et~al.}(2004)\citenamefont{Bienfang, Gross,
  Mink, Hershman, Nakassis, Tang, Lu, Su, Clark, Williams
  et~al.}}]{bienfang2004qkd}
\bibinfo{author}{\bibfnamefont{J.}~\bibnamefont{Bienfang}},
  \bibinfo{author}{\bibfnamefont{A.}~\bibnamefont{Gross}},
  \bibinfo{author}{\bibfnamefont{A.}~\bibnamefont{Mink}},
  \bibinfo{author}{\bibfnamefont{B.}~\bibnamefont{Hershman}},
  \bibinfo{author}{\bibfnamefont{A.}~\bibnamefont{Nakassis}},
  \bibinfo{author}{\bibfnamefont{X.}~\bibnamefont{Tang}},
  \bibinfo{author}{\bibfnamefont{R.}~\bibnamefont{Lu}},
  \bibinfo{author}{\bibfnamefont{D.}~\bibnamefont{Su}},
  \bibinfo{author}{\bibfnamefont{C.}~\bibnamefont{Clark}},
  \bibinfo{author}{\bibfnamefont{C.}~\bibnamefont{Williams}},
  \bibnamefont{et~al.}, \bibinfo{journal}{Optics Express}
  \textbf{\bibinfo{volume}{12}}, \bibinfo{pages}{2011} (\bibinfo{year}{2004}).

\bibitem[{\citenamefont{Schmitt-Manderbach
  et~al.}(2007)\citenamefont{Schmitt-Manderbach, Weier, F{\"u}rst, Ursin,
  Tiefenbacher, Scheidl, Perdigues, Sodnik, Kurtsiefer, Rarity
  et~al.}}]{Schmitt-Manderbach2007}
\bibinfo{author}{\bibfnamefont{T.}~\bibnamefont{Schmitt-Manderbach}},
  \bibinfo{author}{\bibfnamefont{H.}~\bibnamefont{Weier}},
  \bibinfo{author}{\bibfnamefont{M.}~\bibnamefont{F{\"u}rst}},
  \bibinfo{author}{\bibfnamefont{R.}~\bibnamefont{Ursin}},
  \bibinfo{author}{\bibfnamefont{F.}~\bibnamefont{Tiefenbacher}},
  \bibinfo{author}{\bibfnamefont{T.}~\bibnamefont{Scheidl}},
  \bibinfo{author}{\bibfnamefont{J.}~\bibnamefont{Perdigues}},
  \bibinfo{author}{\bibfnamefont{Z.}~\bibnamefont{Sodnik}},
  \bibinfo{author}{\bibfnamefont{C.}~\bibnamefont{Kurtsiefer}},
  \bibinfo{author}{\bibfnamefont{J.}~\bibnamefont{Rarity}},
  \bibnamefont{et~al.}, \bibinfo{journal}{Physical Review Letters}
  \textbf{\bibinfo{volume}{98}}, \bibinfo{pages}{10504} (\bibinfo{year}{2007}).

\bibitem[{\citenamefont{Aspelmeyer
  et~al.}(2003{\natexlab{b}})\citenamefont{Aspelmeyer, B\"ohm, Gyatso,
  Jennewein, Kaltenbaek, Lindenthal, Molina-Terriza, Poppe, Resch, Taraba
  et~al.}}]{aspelmeyer2003ldf}
\bibinfo{author}{\bibfnamefont{M.}~\bibnamefont{Aspelmeyer}},
  \bibinfo{author}{\bibfnamefont{H.~R.} \bibnamefont{B\"ohm}},
  \bibinfo{author}{\bibfnamefont{T.}~\bibnamefont{Gyatso}},
  \bibinfo{author}{\bibfnamefont{T.}~\bibnamefont{Jennewein}},
  \bibinfo{author}{\bibfnamefont{R.}~\bibnamefont{Kaltenbaek}},
  \bibinfo{author}{\bibfnamefont{M.}~\bibnamefont{Lindenthal}},
  \bibinfo{author}{\bibfnamefont{G.}~\bibnamefont{Molina-Terriza}},
  \bibinfo{author}{\bibfnamefont{A.}~\bibnamefont{Poppe}},
  \bibinfo{author}{\bibfnamefont{K.}~\bibnamefont{Resch}},
  \bibinfo{author}{\bibfnamefont{M.}~\bibnamefont{Taraba}},
  \bibnamefont{et~al.}, \bibinfo{journal}{Science}
  \textbf{\bibinfo{volume}{301}}, \bibinfo{pages}{621}
  (\bibinfo{year}{2003}{\natexlab{b}}).

\bibitem[{\citenamefont{Peng et~al.}(2005)\citenamefont{Peng, Yang, Bao, Zhang,
  Jin, Feng, Yang, Yang, Yin, Zhang et~al.}}]{peng2005efs}
\bibinfo{author}{\bibfnamefont{C.~Z.} \bibnamefont{Peng}},
  \bibinfo{author}{\bibfnamefont{T.}~\bibnamefont{Yang}},
  \bibinfo{author}{\bibfnamefont{X.~H.} \bibnamefont{Bao}},
  \bibinfo{author}{\bibfnamefont{J.}~\bibnamefont{Zhang}},
  \bibinfo{author}{\bibfnamefont{X.~M.} \bibnamefont{Jin}},
  \bibinfo{author}{\bibfnamefont{F.~Y.} \bibnamefont{Feng}},
  \bibinfo{author}{\bibfnamefont{B.}~\bibnamefont{Yang}},
  \bibinfo{author}{\bibfnamefont{J.}~\bibnamefont{Yang}},
  \bibinfo{author}{\bibfnamefont{J.}~\bibnamefont{Yin}},
  \bibinfo{author}{\bibfnamefont{Q.}~\bibnamefont{Zhang}},
  \bibnamefont{et~al.}, \bibinfo{journal}{Physical Review Letters}
  \textbf{\bibinfo{volume}{94}}, \bibinfo{pages}{150501}
  (\bibinfo{year}{2005}).

\bibitem[{\citenamefont{Resch et~al.}(2005)\citenamefont{Resch, Lindenthal,
  Blauensteiner, B{\"o}hm, Fedrizzi, Kurtsiefer, Poppe, Schmitt-Manderbach,
  Taraba, Ursin et~al.}}]{resch2005dea}
\bibinfo{author}{\bibfnamefont{K.}~\bibnamefont{Resch}},
  \bibinfo{author}{\bibfnamefont{M.}~\bibnamefont{Lindenthal}},
  \bibinfo{author}{\bibfnamefont{B.}~\bibnamefont{Blauensteiner}},
  \bibinfo{author}{\bibfnamefont{H.}~\bibnamefont{B{\"o}hm}},
  \bibinfo{author}{\bibfnamefont{A.}~\bibnamefont{Fedrizzi}},
  \bibinfo{author}{\bibfnamefont{C.}~\bibnamefont{Kurtsiefer}},
  \bibinfo{author}{\bibfnamefont{A.}~\bibnamefont{Poppe}},
  \bibinfo{author}{\bibfnamefont{T.}~\bibnamefont{Schmitt-Manderbach}},
  \bibinfo{author}{\bibfnamefont{M.}~\bibnamefont{Taraba}},
  \bibinfo{author}{\bibfnamefont{R.}~\bibnamefont{Ursin}},
  \bibnamefont{et~al.}, \bibinfo{journal}{Optics Express}
  \textbf{\bibinfo{volume}{13}}, \bibinfo{pages}{202} (\bibinfo{year}{2005}).

\bibitem[{\citenamefont{Marcikic et~al.}(2006)\citenamefont{Marcikic,
  Lamas-Linares, and Kurtsiefer}}]{marcikic2006fsq}
\bibinfo{author}{\bibfnamefont{I.}~\bibnamefont{Marcikic}},
  \bibinfo{author}{\bibfnamefont{A.}~\bibnamefont{Lamas-Linares}},
  \bibnamefont{and}
  \bibinfo{author}{\bibfnamefont{C.}~\bibnamefont{Kurtsiefer}},
  \bibinfo{journal}{Applied Physics Letters} \textbf{\bibinfo{volume}{89}},
  \bibinfo{pages}{101122} (\bibinfo{year}{2006}).

\bibitem[{\citenamefont{Ursin et~al.}(2007)\citenamefont{Ursin, Tiefenbacher,
  Schmitt-Manderbach, Weier, Scheidl, Lindenthal, Blauensteiner, Jennewein,
  Perdigues, Trojek et~al.}}]{ursin2007ebq}
\bibinfo{author}{\bibfnamefont{R.}~\bibnamefont{Ursin}},
  \bibinfo{author}{\bibfnamefont{F.}~\bibnamefont{Tiefenbacher}},
  \bibinfo{author}{\bibfnamefont{T.}~\bibnamefont{Schmitt-Manderbach}},
  \bibinfo{author}{\bibfnamefont{H.}~\bibnamefont{Weier}},
  \bibinfo{author}{\bibfnamefont{T.}~\bibnamefont{Scheidl}},
  \bibinfo{author}{\bibfnamefont{M.}~\bibnamefont{Lindenthal}},
  \bibinfo{author}{\bibfnamefont{B.}~\bibnamefont{Blauensteiner}},
  \bibinfo{author}{\bibfnamefont{T.}~\bibnamefont{Jennewein}},
  \bibinfo{author}{\bibfnamefont{J.}~\bibnamefont{Perdigues}},
  \bibinfo{author}{\bibfnamefont{P.}~\bibnamefont{Trojek}},
  \bibnamefont{et~al.}, \bibinfo{journal}{Nature Physics}
  \textbf{\bibinfo{volume}{3}}, \bibinfo{pages}{481} (\bibinfo{year}{2007}).

\bibitem[{\citenamefont{Clauser et~al.}(1969)\citenamefont{Clauser, Horne,
  Shimony, and Holt}}]{clauser1969pet}
\bibinfo{author}{\bibfnamefont{J.~F.} \bibnamefont{Clauser}},
  \bibinfo{author}{\bibfnamefont{M.~A.} \bibnamefont{Horne}},
  \bibinfo{author}{\bibfnamefont{A.}~\bibnamefont{Shimony}}, \bibnamefont{and}
  \bibinfo{author}{\bibfnamefont{R.~A.} \bibnamefont{Holt}},
  \bibinfo{journal}{Physical Review Letters} \textbf{\bibinfo{volume}{23}},
  \bibinfo{pages}{880} (\bibinfo{year}{1969}).

\bibitem[{\citenamefont{Mattle et~al.}(1996)\citenamefont{Mattle, Weinfurter,
  Kwiat, and Zeilinger}}]{mattle1996}
\bibinfo{author}{\bibfnamefont{K.}~\bibnamefont{Mattle}},
  \bibinfo{author}{\bibfnamefont{H.}~\bibnamefont{Weinfurter}},
  \bibinfo{author}{\bibfnamefont{P.~G.} \bibnamefont{Kwiat}}, \bibnamefont{and}
  \bibinfo{author}{\bibfnamefont{A.}~\bibnamefont{Zeilinger}},
  \bibinfo{journal}{Physical Review Letters} \textbf{\bibinfo{volume}{76}},
  \bibinfo{pages}{4656} (\bibinfo{year}{1996}).

\bibitem[{\citenamefont{Pan et~al.}(2003)\citenamefont{Pan, Gasparoni, Ursin,
  Weihs, and Zeilinger}}]{pan2003eep}
\bibinfo{author}{\bibfnamefont{J.~W.} \bibnamefont{Pan}},
  \bibinfo{author}{\bibfnamefont{S.}~\bibnamefont{Gasparoni}},
  \bibinfo{author}{\bibfnamefont{R.}~\bibnamefont{Ursin}},
  \bibinfo{author}{\bibfnamefont{G.}~\bibnamefont{Weihs}}, \bibnamefont{and}
  \bibinfo{author}{\bibfnamefont{A.}~\bibnamefont{Zeilinger}},
  \bibinfo{journal}{Nature} \textbf{\bibinfo{volume}{423}},
  \bibinfo{pages}{417} (\bibinfo{year}{2003}).

\bibitem[{\citenamefont{Bouwmeester et~al.}(2001)\citenamefont{Bouwmeester,
  Ekert, and Zeilinger}}]{bouwmeester2001pqi}
\bibinfo{author}{\bibfnamefont{D.}~\bibnamefont{Bouwmeester}},
  \bibinfo{author}{\bibfnamefont{A.~K.} \bibnamefont{Ekert}}, \bibnamefont{and}
  \bibinfo{author}{\bibfnamefont{A.}~\bibnamefont{Zeilinger}},
  \emph{\bibinfo{title}{{The Physics of Quantum Information: Quantum
  Cryptography, Quantum Teleportation, Quantum Computation}}}
  (\bibinfo{publisher}{Springer}, \bibinfo{year}{2001}).

\bibitem[{\citenamefont{Boileau et~al.}(2004)\citenamefont{Boileau, Gottesman,
  Laflamme, Poulin, and Spekkens}}]{boileau2004rpb}
\bibinfo{author}{\bibfnamefont{J.}~\bibnamefont{Boileau}},
  \bibinfo{author}{\bibfnamefont{D.}~\bibnamefont{Gottesman}},
  \bibinfo{author}{\bibfnamefont{R.}~\bibnamefont{Laflamme}},
  \bibinfo{author}{\bibfnamefont{D.}~\bibnamefont{Poulin}}, \bibnamefont{and}
  \bibinfo{author}{\bibfnamefont{R.}~\bibnamefont{Spekkens}},
  \bibinfo{journal}{Physical Review Letters} \textbf{\bibinfo{volume}{92}},
  \bibinfo{pages}{17901} (\bibinfo{year}{2004}).

\bibitem[{\citenamefont{Ma et~al.}(2007)\citenamefont{Ma, Fung, and
  Lo}}]{ma2007qkd}
\bibinfo{author}{\bibfnamefont{X.}~\bibnamefont{Ma}},
  \bibinfo{author}{\bibfnamefont{C.}~\bibnamefont{Fung}}, \bibnamefont{and}
  \bibinfo{author}{\bibfnamefont{H.}~\bibnamefont{Lo}},
  \bibinfo{journal}{Physical Review A} \textbf{\bibinfo{volume}{76}},
  \bibinfo{pages}{12307} (\bibinfo{year}{2007}).

\bibitem[{\citenamefont{Kim et~al.}(2006)\citenamefont{Kim, Fiorentino, and
  Wong}}]{kim2006pss}
\bibinfo{author}{\bibfnamefont{T.}~\bibnamefont{Kim}},
  \bibinfo{author}{\bibfnamefont{M.}~\bibnamefont{Fiorentino}},
  \bibnamefont{and} \bibinfo{author}{\bibfnamefont{F.}~\bibnamefont{Wong}},
  \bibinfo{journal}{Physical Review A} \textbf{\bibinfo{volume}{73}},
  \bibinfo{pages}{12316} (\bibinfo{year}{2006}).

\bibitem[{\citenamefont{Fedrizzi et~al.}(2007)\citenamefont{Fedrizzi, Herbst,
  Poppe, Jennewein, and Zeilinger}}]{fedrizzi2007wtf}
\bibinfo{author}{\bibfnamefont{A.}~\bibnamefont{Fedrizzi}},
  \bibinfo{author}{\bibfnamefont{T.}~\bibnamefont{Herbst}},
  \bibinfo{author}{\bibfnamefont{A.}~\bibnamefont{Poppe}},
  \bibinfo{author}{\bibfnamefont{T.}~\bibnamefont{Jennewein}},
  \bibnamefont{and}
  \bibinfo{author}{\bibfnamefont{A.}~\bibnamefont{Zeilinger}},
  \bibinfo{journal}{Opt. Express} \textbf{\bibinfo{volume}{15}},
  \bibinfo{pages}{15377} (\bibinfo{year}{2007}).

\bibitem[{\citenamefont{Bennett et~al.}(1992)\citenamefont{Bennett, Brassard,
  and Mermin}}]{bennett1992qcw}
\bibinfo{author}{\bibfnamefont{C.~H.} \bibnamefont{Bennett}},
  \bibinfo{author}{\bibfnamefont{G.}~\bibnamefont{Brassard}}, \bibnamefont{and}
  \bibinfo{author}{\bibfnamefont{N.~D.} \bibnamefont{Mermin}},
  \bibinfo{journal}{Physical Review Letters} \textbf{\bibinfo{volume}{68}},
  \bibinfo{pages}{557} (\bibinfo{year}{1992}).

\bibitem[{\citenamefont{Perdigues-Armengol
  et~al.}(2007)\citenamefont{Perdigues-Armengol, Furch, de~Matos, Minster,
  Cacciapuoti, Pfennigbauer, Aspelmeyer, Jennewein, Ursin, Schmitt-Manderbach
  et~al.}}]{perdigues2007qce}
\bibinfo{author}{\bibfnamefont{J.~M.} \bibnamefont{Perdigues-Armengol}},
  \bibinfo{author}{\bibfnamefont{B.}~\bibnamefont{Furch}},
  \bibinfo{author}{\bibfnamefont{C.~J.} \bibnamefont{de~Matos}},
  \bibinfo{author}{\bibfnamefont{O.}~\bibnamefont{Minster}},
  \bibinfo{author}{\bibfnamefont{L.}~\bibnamefont{Cacciapuoti}},
  \bibinfo{author}{\bibfnamefont{M.}~\bibnamefont{Pfennigbauer}},
  \bibinfo{author}{\bibfnamefont{M.}~\bibnamefont{Aspelmeyer}},
  \bibinfo{author}{\bibfnamefont{T.}~\bibnamefont{Jennewein}},
  \bibinfo{author}{\bibfnamefont{R.}~\bibnamefont{Ursin}},
  \bibinfo{author}{\bibfnamefont{T.}~\bibnamefont{Schmitt-Manderbach}},
  \bibnamefont{et~al.}, \bibinfo{journal}{58th International Astronautical
  Congress, Hyderabad, India}  (\bibinfo{year}{2007}).

\end{thebibliography}

\end{document}